\documentclass[aps,twocolumn,superscriptaddress,longbibliography]{revtex4-2}

\usepackage{amsmath}
\usepackage{amssymb}
\usepackage{amstext}
\usepackage{amsopn}
\usepackage{amsfonts}
\usepackage{amsxtra}
\usepackage[english]{babel}
\usepackage{graphicx}
\usepackage{bm}
\usepackage{multirow}
\usepackage{dcolumn}
\usepackage{color}
\usepackage{verbatim}
\def\AVS{AV$_{3}$Sb$_{5}$}
\def\CVS{CsV$_{3}$Sb$_{5}$}
\def\RVS{RbV$_{3}$Sb$_{5}$}
\def\KVS{KV$_{3}$Sb$_{5}$}

\def\mathbi#1{\ensuremath{\textbf{\em #1}}}

\def\QCDW{\ensuremath{\mathbf{Q}_{\text{CDW}}}}
\def\TCDW{\ensuremath{T_{\text{CDW}}}}

\begin{document}

\title{Observation of Unconventional Charge Density Wave without Acoustic Phonon Anomaly in Kagome Superconductors \AVS{} (A=Rb,Cs)}

\author{H. X. Li}
\affiliation{Material Science and Technology Division, Oak Ridge National Laboratory, Oak Ridge, Tennessee 37831, USA}
\author{T. T. Zhang}
\affiliation{Department of Physics, Tokyo Institute of Technology, Okayama, Meguro-ku, Tokyo 152-8551, Japan}
\affiliation{Tokodai Institute for Element Strategy, Tokyo Institute of Technology, Nagatsuta, Midori-ku, Yokohama, Kanagawa 226-8503, Japan}
\author{T. Yilmaz}
\affiliation{National Synchrotron Light Source II, Brookhaven National Laboratory, Upton, New York 11973, USA}
\author{Y. Y. Pai}
\affiliation{Material Science and Technology Division, Oak Ridge National Laboratory, Oak Ridge, Tennessee 37831, USA}
\author{C. Marvinney}
\affiliation{Material Science and Technology Division, Oak Ridge National Laboratory, Oak Ridge, Tennessee 37831, USA}
\author{A. Said}
\affiliation{Advanced Photon Source, Argonne National Laboratory, Argonne, Illinois 60439, USA}
\author{Q. Yin}
\affiliation{Department of Physics and Beijing Key Laboratory of Opto-Electronic Functional Materials and Micro-devices, Renmin University of China, Beijing, China}
\author{C. Gong}
\affiliation{Department of Physics and Beijing Key Laboratory of Opto-Electronic Functional Materials and Micro-devices, Renmin University of China, Beijing, China}
\author{Z. Tu}
\affiliation{Department of Physics and Beijing Key Laboratory of Opto-Electronic Functional Materials and Micro-devices, Renmin University of China, Beijing, China}
\author{E. Vescovo}
\affiliation{National Synchrotron Light Source II, Brookhaven National Laboratory, Upton, New York 11973, USA}
\author{R. G. Moore}
\affiliation{Material Science and Technology Division, Oak Ridge National Laboratory, Oak Ridge, Tennessee 37831, USA}
\author{S. Murakami}
\affiliation{Department of Physics, Tokyo Institute of Technology, Okayama, Meguro-ku, Tokyo 152-8551, Japan}
\affiliation{Tokodai Institute for Element Strategy, Tokyo Institute of Technology, Nagatsuta, Midori-ku, Yokohama, Kanagawa 226-8503, Japan}
\author{H. C. Lei}\email{hlei@ruc.edu.cn}
\affiliation{Department of Physics and Beijing Key Laboratory of Opto-Electronic Functional Materials and Micro-devices, Renmin University of China, Beijing, China}
\author{H. N. Lee}
\affiliation{Material Science and Technology Division, Oak Ridge National Laboratory, Oak Ridge, Tennessee 37831, USA}
\author{B. Lawrie}
\affiliation{Material Science and Technology Division, Oak Ridge National Laboratory, Oak Ridge, Tennessee 37831, USA}
\author{H. Miao} \email{miaoh@ornl.gov}
\affiliation{Material Science and Technology Division, Oak Ridge National Laboratory, Oak Ridge, Tennessee 37831, USA}

\date{\today}

\begin{abstract}
The combination of non-trivial band topology and symmetry breaking phases gives rise to novel quantum states and phenomena such as topological superconductivity, quantum anomalous Hall effect and axion electrodynamics. Evidence of intertwined charge density wave (CDW) and superconducting order parameters has recently been observed in a novel kagome material \AVS{} (A=K,Rb,Cs) that features a $\mathbb{Z}_2$ topological invariant in the electronic structure. However, the origin of the CDW and its intricate interplay with topological state has yet to be determined. Here, using hard x-ray scattering, we demonstrate a three-dimensional (3D) CDW with $2\times2\times2$ superstructure in (Rb,Cs)V$_3$Sb$_5$. Unexpectedly, we find that the CDW fails to induce acoustic phonon anomalies at the CDW wavevector but yields a novel Raman mode that quickly damps into a broad continuum below the CDW transition temperature. Our observations exclude strong electron-phonon coupling driven CDW in \AVS{} and point to an unconventional particle-hole condensation mechanism that couples CDW, superconductivity and topological band structure.
\end{abstract}
\maketitle
\begin{figure*}
\includegraphics[width=0.8\textwidth]{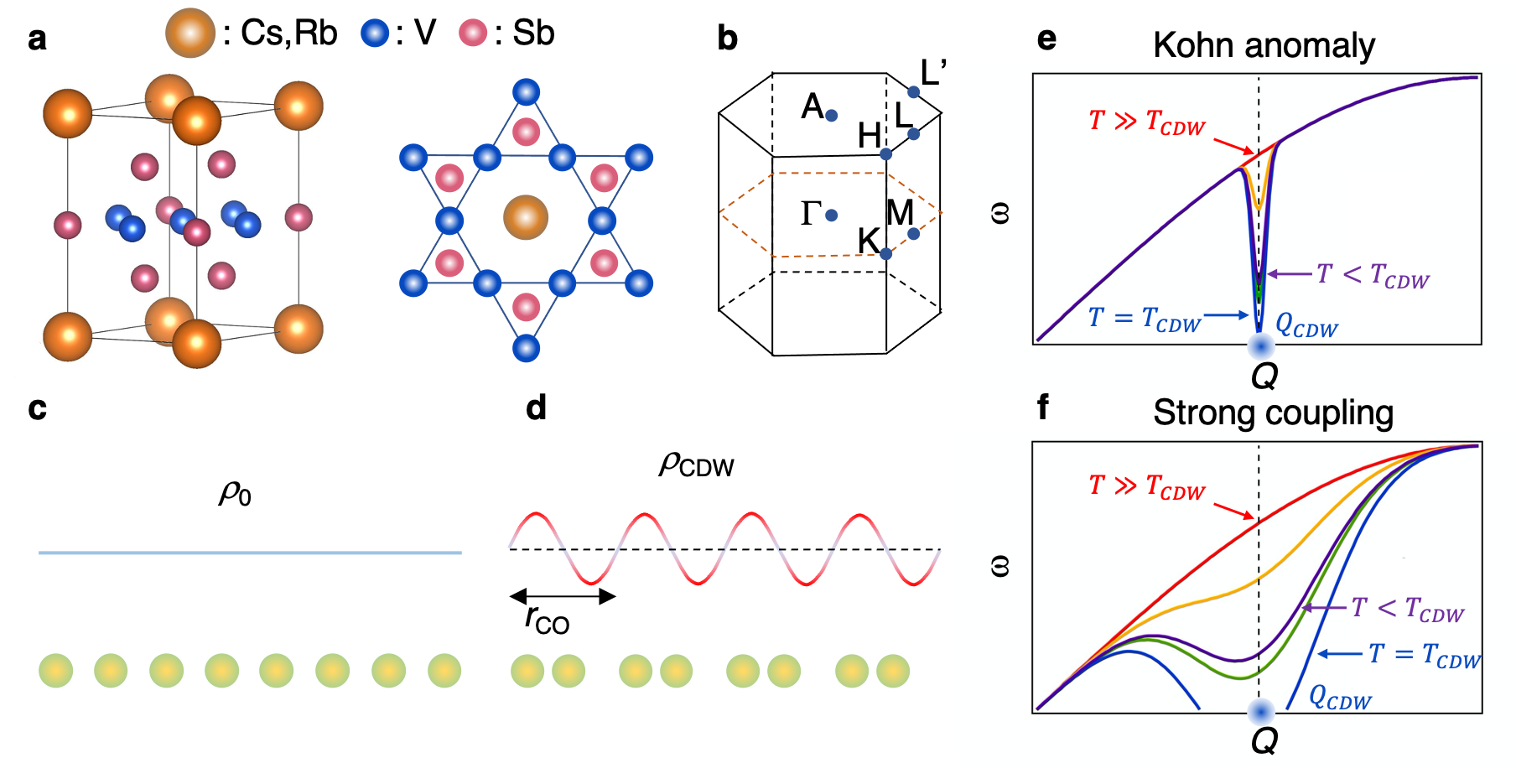}
\caption{The crystal structure of \AVS{} and CDW induced phonon anomalies. (a) shows the crystal structure of \AVS{} (space group P6/mmm, No. 191), which consists of V-Sb slabs that are separated by alkali elements. The V-Sb slab contains a V-kagome lattice and two Sb sites laying in the kagome plane and above the V-triangle. (b) shows the 3D BZ and high symmetry points of \AVS{}. (c) and (d) show the valence and core electron density in the normal and CDW phase, respectively. Due to EPC, the core electrons will follow the CDW period and yield acoustic phonon anomalies. (e) Nesting induced Kohn anomalies.  (f) Strong coupling induced phonon softening in an extended momentum space near \QCDW. Red, yellow, green, blue and purple colors represent temperature from $T\gg$ \TCDW{} to $T<$\TCDW. The phonon hardening for $T<$\TCDW\ is due to the CDW amplitude mode.\label{Fig1}}

\end{figure*}

The charge density wave (CDW), a translational symmetry-breaking electronic fluid, plays a crucial role in unconventional superconductors and intertwined electronic orders~\cite{Fradkin2015,Lee2014,Miao2019,Thomale2013,Wang2013}. While CDWs have been isolated from topological excitations, recently an experimental evidence of topological CDW with chiral flux is observed in a new kagome metal, \AVS{} (A=K, Rb, Cs)~\cite{jiang2020discovery}, whose crystal structure and 3D Brillouin zone are shown in Fig.~\ref{Fig1}(a) and (b), respectively.
%
Density functional theory (DFT) calculations of the electronic structure find a $\mathbb{Z}_2$ topological invariant and saddle points at the M point that may lead to a CDW instability at low temperature~\cite{Ortiz2019, Ortiz2020,jiang2020discovery,tan2021charge,feng2021chiral}. Interestingly, scanning tunneling microscopy (STM) studies of \CVS{} observed evidence of roton pair-density wave and Majorana zero mode, suggesting novel topological superconductivity intertwines with CDW~\cite{liang2021threedimensional,chen2021roton}. Despite the intimate correlations between the CDW, superconductivity (SC) and topological band structure~\cite{Ortiz2019, Ortiz2020,jiang2020discovery,Yin2021,tan2021charge,feng2021chiral,zhao2021cascade,liang2021threedimensional,chen2021roton,yu2021concurrence,chen2021double}, the nature of the CDW, in particular the role of electron-phonon coupling (EPC), remains unresolved, hindering the proper interpretations of SC in \AVS{}.

Theoretically, the CDW mechanisms are broadly separated into two categories: weak coupling scenarios based on Fermi-surface instabilities~\cite{Mazin2008,Rice1975,Hoesch2009} and strong coupling theories derived from local electron-electron or electron-phonon interactions~\cite{Zaanen1989,MACHIDA1989192,Poilblanc1989,Varma1983}. As we show in Fig.~\ref{Fig1}(c) and (d), due to the finite EPC, the formation of CDWs distorts the underlying lattice and typically results in acoustic phonon anomalies. For a weak coupling mechanism, the phonon softening is sharply confined near the CDW wavevector,~\QCDW, and is known as Kohn anomaly\cite{Hoesch2009} [illustrated in Fig.~\ref{Fig1}(e)]. In contrast, strong coupling mechanisms usually give rise to phonon anomalies in an extended \textbf{Q}-range near~\QCDW{} \cite{Miao2018,LeTacon2014,Weber2011,Kogar1314}, as shown in Fig.~\ref{Fig1}(f). Here we combine inelastic x-ray scattering (IXS), Raman spectroscopy and angle-resolved photoemission spectroscopy (ARPES) to uncover a novel particle-hole condensation in \AVS{} (A=Cs, Rb). Unlike conventional CDW materials, we demonstrate that neither longitudinal (LA) nor transverse (TA) acoustic phonon modes show CDW induced phonon anomalies  in \AVS{} (A=Cs, Rb). Instead, we discover a novel Raman mode that emerges at the CDW temperature, \TCDW, and is quickly damped below \TCDW{} into a continuum that broadly peaked at 20~meV. Intriguingly, the three-dimensional (3D) $2\times2\times2$ superstructure, determined by hard x-ray diffraction (XRD), connects particle-hole scattering channels in the electronic band structure and yields a  band-inversion near the Fermi level, $E_\text{F}$, below \TCDW. Our results uncover the intricate interplay between non-trivial band topology and intertwined symmetry breaking phases and shed light on the unusual SC in \CVS{}. 

\begin{figure*}
\includegraphics[width=0.8\textwidth]{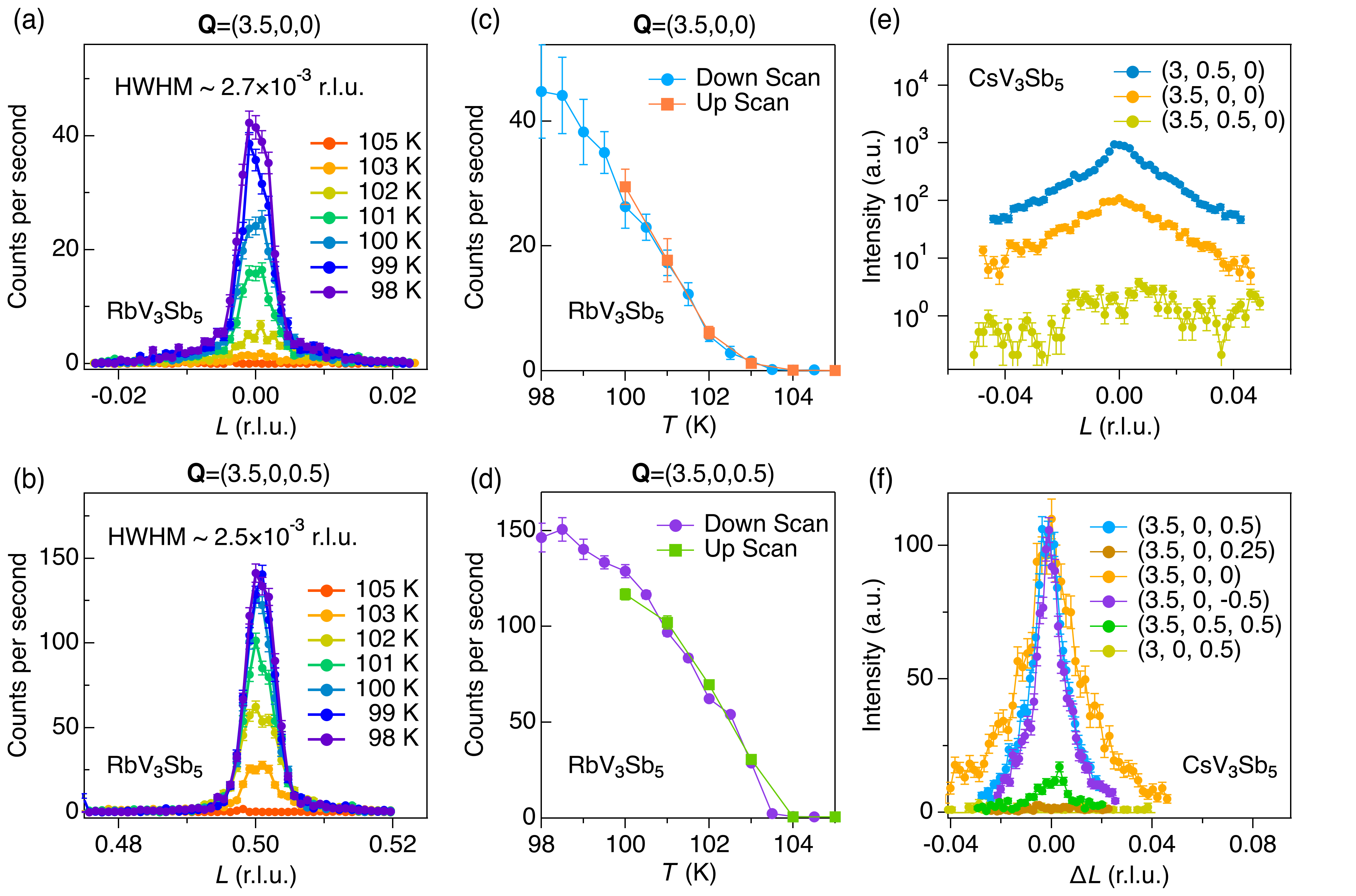}
\caption{3D-CDW determined by meV-resolution XRD. (a,b) The temperature dependent CDW peaks of \RVS{} at $\mathbf{Q}=(3.5, 0, 0)$ and $(3.5, 0, 0.5)$. The CDW peak at half-integer $L$ demonstrates a 3D CDW with $2\times2\times2$  superstructure. The CDW peaks are extremely sharp along $L$, with HWHM of $\sim 0.0025$~r.l.u., at T=98~K. (c) and (d) show the extracted temperature dependent CDW peak intensity at $\mathbf{Q}=(3.5, 0, 0)$ and $(3.5, 0, 0.5)$, respectively. The peak intensity takes a sharp upturn at the CDW onset temperature of 103~K for \RVS, which is consistent with the specific heat measurement (see Appendix A). The absence of hysteresis is consistent with a second order phase transition. CDW peaks in \CVS{} are shown in (e) and (f). The peak intensity at $\mathbf{Q}=(3, 0.5, 0)$ is over 10 times larger than the one at $\mathbf{Q}= (3.5,0,0)$ [panel (e)], consistent with larger structure factors at $\mathbf{Q}=(H, K+0.5, 0)$, $H=$odd integer~\cite{Ortiz2020}. The absence of peaks at $\mathbf{Q}=(3, 0, 0.5)$, $(3.5, 0.5, 0)$, $(3.5, 0, 0.25)$ confirms the same $2\times2\times2$  superstructure in \CVS. The error bars in panel (a),(b),(e),(f) represent one standard deviation assuming Poisson counting statistics. The error bars in panel (c) and (d) denote the $2\sigma$ returned from the pseudo-voigt fittings that extract the peak intensity.\label{Fig2}}

\end{figure*}

We first determine the bulk CDW superlattice peaks using meV resolution XRD. Figure ~\ref{Fig2}(a) and (b) show the temperature dependent $L$-scan of \RVS{} at $\mathbf{Q}=(3.5, 0, 0)$ and $(3.5, 0, 0.5)$ in reciprocal lattice units (r.l.u.). 
%
The onset temperature at $T=103$~K is consistent with the CDW transition temperature in \RVS{} (see Appendix A). The absence of hysteresis, as shown in Fig.~\ref{Fig2}(c) and (d), suggests that the CDW order is a second order phase transition. The observation of a CDW peak at half integer $L$ in Fig.~\ref{Fig2}(b) proves that the CDW in \RVS{} forms a 3D $2\times2\times2$ superstructure rather than a 2D $2\times2$ superstructure as suggested by previous STM and XRD measurements~\cite{Ortiz2020,jiang2020discovery,zhao2021cascade,liang2021threedimensional,chen2021roton}. Indeed, the CDW correlation length, determined by the half-width-at-half-maximum (HWHM), is larger than 1000 {\AA} along the $L$-direction, establishing a long range ordered 3D-CDW. As shown in Fig.~\ref{Fig2}(e) and (f), the same 3D-CDW superlattice structure is also observed in \CVS, suggesting a ubiquitous CDW mechanism for all \AVS. Interestingly, we find that while the CDW is long-range ordered, the integrated CDW intensity that is proportional to the CDW order parameter is extremely small. Comparing with fundamental Bragg peaks, the CDW peak intensity is $3\sim 5$ orders of magnitude weaker, demonstrating small lattice distortions. 

\begin{figure*}
\includegraphics[width=1\textwidth]{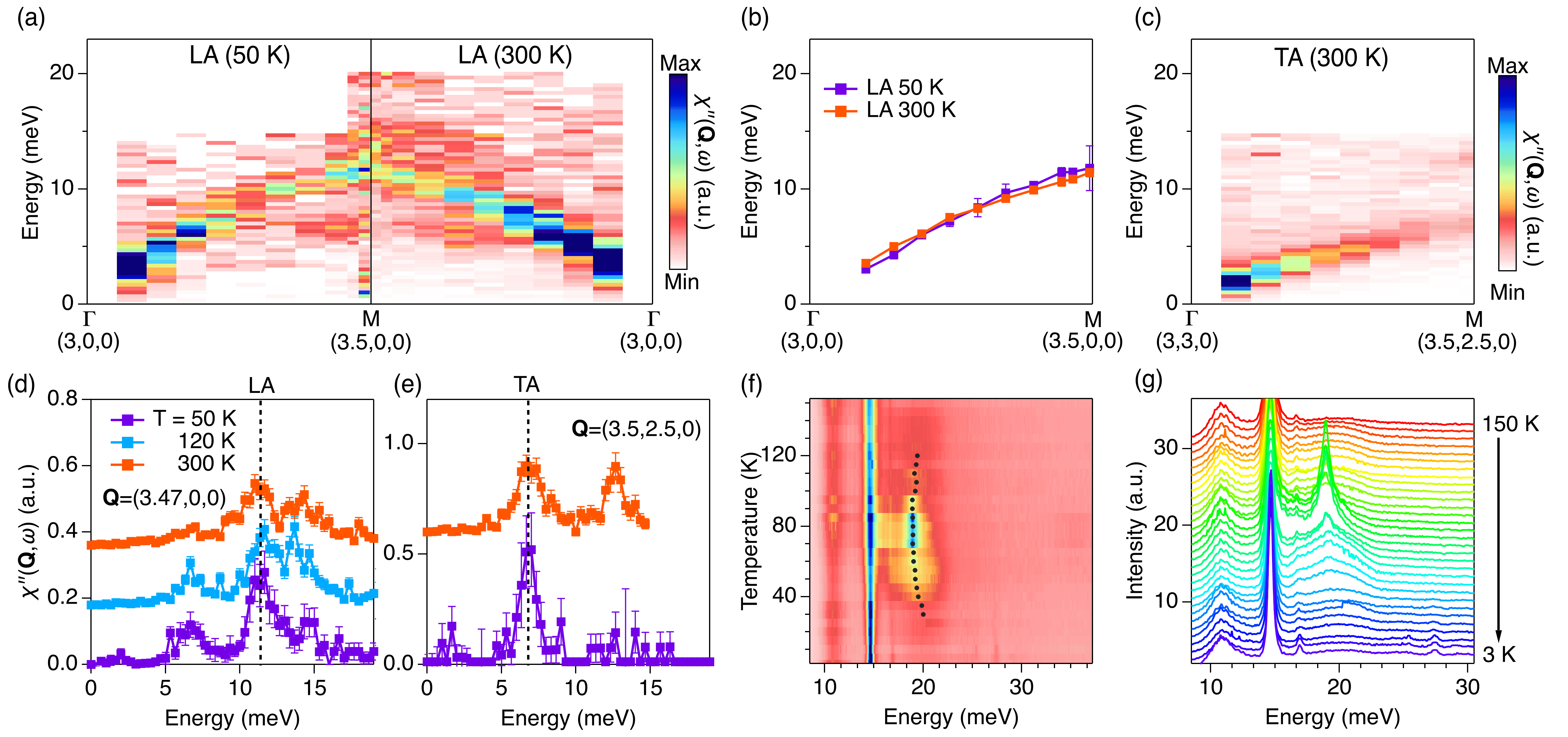}
\caption{Evidence of unconventional electronically driven CDW. (a) compares the IXS determined LA of \RVS{} along $\Gamma(3, 0, 0)$ to M$(3.5, 0, 0)$ at temperatures of $T=300$~K and 50~K (\TCDW=103~K). (b) Extracted temperature dependent phonon dispersions from (a). (c) TA in \RVS{} measured along $\Gamma(3, 3, 0)$ to M$(3.5, 2.5, 0)$ at room temperature (300~K). (d) and (e) compare temperature dependent IXS spectra at $\mathbf{Q}=(3.47, 0, 0)$ and $(3.5, 2.5,0)$. The dashed lines in (d) and (e) highlight LA and TA, respectively. All IXS data are Bose-factor corrected. The raw IXS data is shown in ref.\cite{SI}. (f),(g) Temperature dependent Raman spectra on \CVS{} from 150~K to 3~K. Beside the sharp optical phonon peak at 11, 15 and 17~meV, there is a continuum that is broadly centered near 19~meV (The 11~meV peak may actually be at lower energy but appeared to be at 11~meV due to the cutoff of the longpass filter at 90~cm$^{-1}$). The black dots in (f) mark the peak positions of the continuum near 19~meV. When cooling towards \TCDW, the center of this continuum gradually shifts from 19.5 to 19~meV and starts to build a strong, sharp and asymmetric peak at \TCDW{} = 94~K. Moving to lower temperature, this peak is quickly damped and hardened to 20~meV at 30~K. Below 30~K, two new phonon modes at 25.4 and 27.5~meV ared observed, which may correspond to the stripe phase observed by a recent STM study~\cite{zhao2021cascade}. The error bars in panel (d), (e) represent one standard deviation assuming Poisson counting statistics. The error bars in panel (b) denote the $2\sigma$ returned from the fitting that extract the spectral peak position.\label{Fig3}}

\end{figure*}

With the 3D-CDW established, we explore the origin of the CDW in \AVS{} using IXS and Raman spectroscopy. As we have shown in Fig.~\ref{Fig1}, the formation of CDWs is always accompanied by acoustic phonon anomalies because of EPC. Indeed, even in the cuprates, where CDWs are short ranged and possibly dynamical, a $15\%$ LA phonon softening has been observed~\cite{Miao2018,LeTacon2014}, demonstrating acoustic phonon is a sensitive probe of CDW correlations. Recent DFT calculations of the phonon spectrum of \AVS{} show large negative frequency of the LA near the M and L point, suggesting EPC induced CDW~\cite{Yin2021} (see supplementary materials in ref.\cite{SI}). To test this scenario, we performed IXS measurement along $(3,0,0)-(3.5,0,0)$ and $(3,3,0)-(3.5,2.5,0)$ to selectively enhance LA and TA phonon modes in our experimental geometry (see~\cite{SI}. Unexpectedly, as we show in Fig.~\ref{Fig3}(a)-(e), CDWs in \AVS{} failed to induce phonon anomalies on both TA and LA modes from $T=300$~K~$>$~\TCDW{} to 50~K~$<$~\TCDW. 
%
The absence of phonon softening excludes strong EPC as the driven force of the CDW in \AVS. Instead, our inelastic and elastic results indicate an electronic-driven CDW mechanism with extremely weak electron acoustic-phonon interactions. 

To confirm this scenario, we performed temperature dependent Raman spectroscopy on \CVS{} with \TCDW=94~K. The results are shown in Fig.~\ref{Fig3}(f) and (g). According to DFT calculations (see supplementary materials in~\cite{SI}), four Raman active modes at 8.6, 15.1, 16.2 and 17.4~meV, corresponding to $E_{1g}$, $E_{2g}$, $A_{1g}$ and $B_{2g}$ modes, respectively, are allowed for the normal state structure. At 150~K, we observe three optical phonon modes at 11, 15 and 17~meV, which we attribute to $E_{1g}$, $E_{2g}$ and $B_{2g}$ modes (The 11~meV peak may actually be at lower energy but appeared to be at 11~meV due to the cutoff of the longpass filter at 11~meV). Interestingly, we observe a continuum that is broadly centered around 19.5~meV. As we cool down towards the \TCDW, the center of this continuum gradually shifts to lower energy and eventually develops into a strong, sharp and asymmetric peak at 19~meV, demonstrating its intimate correlation with the CDW. Furthermore, 19~meV is nearly two times of the STM determined single-particle gap at $E_\text{F}$~\cite{jiang2020discovery,zhao2021cascade,liang2021threedimensional,chen2021roton}, consistent with the two-particle process of Raman scattering. Unexpectedly, moving to lower temperature, this peak is suppressed and hardened to 20~meV at 30~K. The novel temperature dependence of the broad Raman mode strongly supports an electronic origin of the CDW. Remarkably, although Raman spectroscopy measures excitations at the Brillouin zone center, this unconventional mode shows clear softening above \TCDW{} and hardening below \TCDW, reminiscent of the CDW phason and amplitudon excitations near the ~\QCDW{} \cite{Weber2011,Kogar1314}. A similar effect has been observed in CDW material TaSe$_2$~\cite{Kogar1314}, which also has a 3D $2\times2\times2$ superstructure. We suspect that CDWs in \AVS{} might be a novel realization of particle-hole condensation in a metallic system, where the particle-hole excitation is highly damped by itinerant electrons. In this scenario, the absence of phonon softening might be a signature of large CDW phason gap that is induced by commensurability effect~\cite{Lee1974}. Indeed, As we discuss in more details in Appendix B, the combination of CDW commensurate period and the unique electronic structure of \AVS{} allows a large gap in the CDW phase mode. We note, below 30~K, we observe additional Raman active modes at 25.4 and 27.5~meV, demonstrating additional symmetry breaking phase at low temperature. Comparing with previous STM study~\cite{zhao2021cascade}, this new low-temperature phase is possibly the unidirectional CDW. Interestingly, this new symmetry breaking phase correlates strongly with the CDW-related continuum, where the intensity of the continuum significantly suppressed when the new Raman modes emerge. 

To disclose the particle-hole scattering channel that is connected by the 3D-CDW, in Fig.~\ref{Fig4}, we show the low-temperature ($T=15$~K) electronic structure of \RVS{} determined by ARPES measurements and DFT calculations. 
%
\begin{figure*}
\includegraphics[width=1\textwidth]{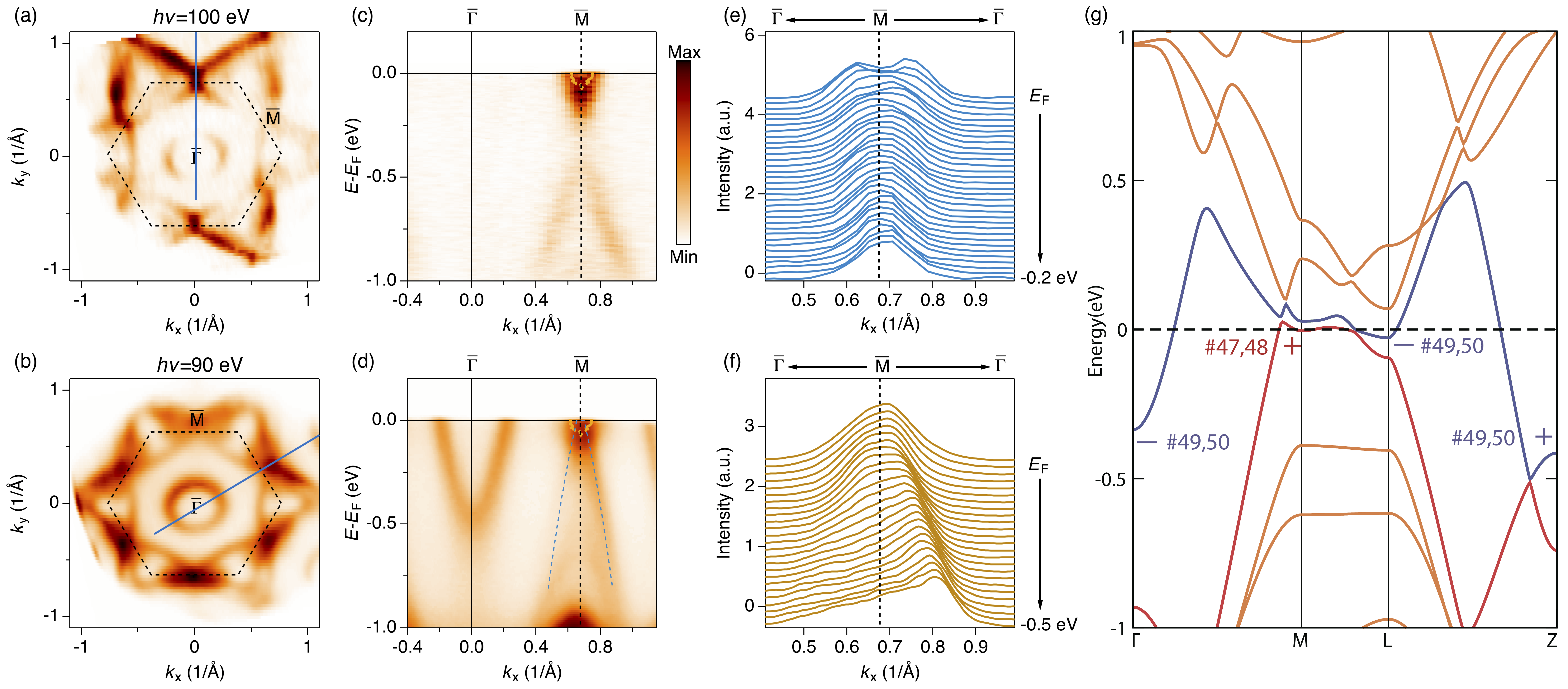}
\caption{3D-CDW induced band-inversion. (a) and (b) show Fermi surface maps measured at photon energies of $h\nu=100$ and 90~eV, corresponding to $k_z=\pi$ and $0.15\pi$ around the $\overline{\text{M}}$ point, respectively. Data was taken at $T=15$~K. The blue lines in (a) and (b) mark the high symmetry cuts shown in (c) and (d), respectively. The inner potential is extracted to be around 8.2~eV (see supplementary materials~\cite{SI}). (c) and (d) selectively enhance the electron band near the unfolded L point and the hole band near the unfolded M point. The yellow dots in (c) and (d) are extracted from the MDC peaks shown in (e). The blue dashed line in (d) is a guide to the eye. The blurry spectral weight near the Fermi level in (d) likely comes from the bulk projected surface state~\cite{Ortiz2020}.(e) and (f) show stacking plots of MDCs to further reveal the electron and hole bands in panel (c) and (d). (g) shows the electronic band structure of \RVS{} calculated by VASP~\cite{Kresse1994,Kresse1996,KRESSE199615} after a full relaxation on atomic positions with an atom pair-wise correction method (DFT-D3)~\cite{Grimme2011} since the van der Waals forces play an important role along the c-layer stacking direction. The Fermi energy is shifted by 100~meV in order to match the experimental results. There is a hole pocket in the vicinity of M point ($\Gamma$-M direction) carrying a parity of ``+'' and an electron pocket in the vicinity of L point carrying a parity of ``-'', contributed by different bands. Our DFT result confirm the parity distribution of the band structure that is reported to possess a non-trivial $\mathbb{Z}_2$ topological invariant and induce topological surface states around M point~\cite{Ortiz2020}. \label{Fig4}}
\end{figure*}
As shown in Fig.~\ref{Fig4}(a) and (b), the overall Fermi surface topology is similar to previous ARPES studies of \KVS{} and \CVS~\cite{Yangeabb6003,Ortiz2020}. Photon energies of $h\nu=100$ and 90~eV are corresponding to $k_z=\pi$ and $0.15\pi$ around the $\overline{\mathrm{M}}$ point, respectively (see supplementary material~\cite{SI} for the $k_z$ map). The most important observations are shown in Fig.~\ref{Fig4}(c) and (d), corresponding to the blue-cut in Fig.~\ref{Fig4}(a) and (b), respectively. At $k_z=\pi$, we observe a shallow electron-band with a band bottom of $\sim 50$~meV, while at $k_z=0.15\pi$, we observe a steep hole-like band dispersing towards the Fermi level. The shallow electron and hole band can be further revealed by the stacking momentum distribution curves (MDCs) displayed in Fig.~\ref{Fig4}(e) and (f). Similar electronic structure is also observed in \CVS{} (see supplementary materials~\cite{SI}). Comparing with the DFT calculations shown in Fig.~\ref{Fig4}(g), we find that the electron and hole bands correspond to band \#49,50 at the L point [Fig.~\ref{Fig4}(g)] and band \#47,48 at the M point, respectively. The parity distribution confirms the opposite parity between these two bands, which is reported to hold a non-trivial $\mathbb{Z}_2$ topology~\cite{Ortiz2020}. We note that at $k_z=0.15\pi$, a weak electron-like spectral weight is observed near the Fermi level. It likely comes from the bulk projected surface state as suggested in a previous study~\cite{Ortiz2020}. Our observations establish two particle-hole channels that are relevant to the CDW: (i) the quasi-nested electron and hole pockets at the M and L point [see Fig.~\ref{Fig1}(b)] and (ii) the saddle points at the M point~\cite{Thomale2013,Wang2013,tan2021charge,feng2021chiral}. The former is likely responsible for the $L$-component of the 3D-CDW , while the latter may play crucial role for the in-plane component as suggested by previous studies~\cite{Thampy2013,Wang2013,feng2021chiral}. Interestingly, in the CDW phase, the shallow electron and hole-like bands will fold to the Brillouin zone center and open a band gap between bands with opposite parity. This CDW induced band inversion near the Fermi level may possibly responsible for the novel SC at low temperature~\cite{zhao2021cascade,liang2021threedimensional,chen2021roton,chen2021double}.

In summary, we demonstrated a novel electronic driven 3D-CDW in the $\mathbb{Z}_2$ kagome superconductor \AVS, where the formation of CDW failed to induce phonon anomalies near ~\QCDW. Our observations exclude strong EPC driven CDW in \AVS{} and point to an unconventional particle-hole condensation mechanism that couples the CDW and the topological band structure.

Acknowledgements: We thank J. G. Cheng, R. Thomale, Z. Q. Wang, S. F. Wu, B. H. Yan and J. X. Yin for stimulating discussions. This research was sponsored by the U.S. Department of Energy, Office of Science, Basic Energy Sciences, Materials Sciences and Engineering Division (IXS, ARPES data analysis, Raman spectroscopy) and by the Laboratory Directed Research and Development Program of Oak Ridge National Laboratory, managed by UT-Battelle, LLC, for the U. S. Department of Energy (IXS and ARPES experiment). This research used resources (IXS experiment at beamline 30ID) of the Advanced Photon Source, a U.S. DOE Office of Science User Facility operated for the DOE Office of Science by Argonne National Laboratory under Contract No. DE-AC02-06CH11357. ARPES measurements used resources at 21-ID-1 beamlines of the National Synchrotron Light Source II, a US Department of Energy Office of Science User Facility operated for the DOE Office of Science by Brookhaven National Laboratory under contract no. DE-SC0012704. Extraordinary facility operations were supported in part by the DOE Office of Science through the National Virtual Biotechnology Laboratory, a consortium of DOE national laboratories focused on the response to COVID-19, with funding provided by the Coronavirus CARES Act. Raman spectroscopy was conducted at the Center for Nanophase Materials Sciences, which is a DOE Office of Science User Facility. T.T. Z. and S. M. acknowledge support from Tokodai Institute for Element Strategy (TIES) funded by MEXT Elements Strategy Initiative to Form Core Research Center. S. M. also acknowledges support by JSPS KAKENHI Grant Number JP18H03678. H.C.L acknowledges the supports from the National Key R\&D Program of China (Grant Nos. 2016YFA0300504, and 2018YFE0202600), the National Natural Science Foundation of China (Grant Nos. 11774423 and 11822412), the Fundamental Research Funds for the Central Universities, and the Research Funds of Renmin University of China (Grant Nos. 18XNLG14 and 19XNLG17).  C.M. was supported by the Intelligence Community Postdoctoral Research Fellowship Program at the Oak Ridge National Laboratory, administered by Oak Ridge Institute for Science and Education through an interagency agreement between the U.S. Department of Energy and the Office of the Director of National Intelligence.
\appendix 

\section{Methods}
\noindent\mathbi{Sample:} Single crystals of \AVS{} (A = Rb, Cs) were grown from Rb ingot (purity 99.9\%), Cs ingot (purity 99.9\%), V powder (purity 99.9\%) and Sb grains (purity 99.999\%) using the self-flux method, similar to the growth of \RVS{}~\cite{Yin2021}. The mixture was put into an alumina crucible and sealed in a quartz ampoule under partial argon atmosphere. The sealed quartz ampoule was heated to 1273~K for 12~h and soaked there for 24~h. Then it was cooled down to 1173~K at 50~K/h and further to 923~K at a slowly rate. Finally, the ampoule was taken out from the furnace and decanted with a centrifuge to separate \CVS{} single crystals from the flux. The obtained crystals have a typical size of $2\times2\times0.02$ mm$^3$. CDW transition is clearly observed in transport and specific heat measurement shown in Fig.~\ref{FigA1}(a) and (b), respectively.

%
\begin{figure}[tb]
\includegraphics[width=8.5 cm]{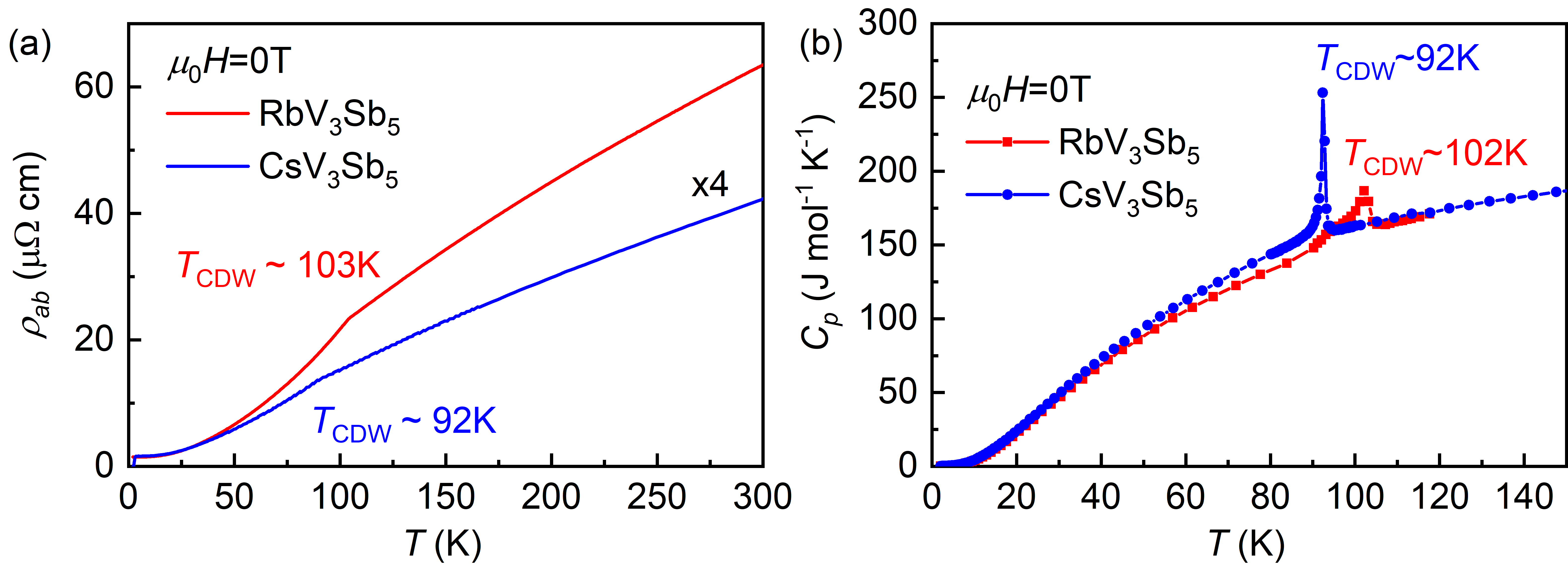}
\renewcommand{\thefigure}{A\arabic{figure}}
\setcounter{figure}{0}
\caption{(a) and (b) show transport and specific heat measurement of \RVS{} (red) and \CVS{}(blue).\label{FigA1}}
\end{figure}

\noindent\mathbi{Inelastic x-ray scattering:} The experiments were conducted at beam line 30-ID-C (HERIX) at the Advanced Photon Source (APS). The highly monochromatic x-ray beam of incident energy $E_i$ = 23.7~keV (l = 0.5226 \AA) was focused on the sample with a beam cross section of 35$\times$15~$\mu$m$^2$ (horizontal × vertical). The total energy resolution of the monochromatic x-ray beam and analyzer crystals was $\Delta E\sim$1.5~meV (full width at half maximum). The measurements were performed in transmission geometry. Typical counting times were in the range of 30 to 360 seconds per point in the energy scans at constant momentum transfer $\mathbf{Q}$. $H,\ K,\ L$ are defined in the hexagonal structure with a=b=5.472~\AA, c=9.073~\AA\ at the room temperature for \RVS{}, and a=b=5.495~\AA, c=9.309~\AA\ at the room temperature for \CVS{}.

\noindent\mathbi{ARPES experiment:} The ARPES experiments were performed on single crystals of \RVS{} and \CVS{}. The samples were cleaved $in$-$situ$ in a vaccum better than 5$\times$10$^{-11}\  torr$. The experiment was performed at beamline 21-ID-1 at National Synchrotron Light Source II, Brookhaven. The measurements were taken with synchrotron light source and a Scienta-Omicron DA30 electron analyzer. The total energy resolution of the ARPES measurement is $\sim$15~meV. The sample stage was maintained at low temperature ($T=$15~K) throughout the experiment.

\noindent\mathbi{Raman experiment:} Raman spectroscopy was performed in a Montana Instruments closed cycle Cryostation s100 and utilized a Hubner Photonics 532~nm diode pumped laser excitation and an Isoplane SCT-320 imaging spectrograph with a 400B-eXcelon CCD camera and a 2400 groove/mm visible-holographic grating. Semrock dichroic and long pass filters were integrated in the optics train. For all Raman spectra reported here, a 2~mW laser excitation power was used, and Raman spectra were acquired for 300~s each.  The sample was initially cooled to 3~K then heated to 150~K. All of the reported Raman spectra were then acquired while monotonically recooling the sample from 150~K to 3~K. The optics train was refocused at each temperature after waiting for the sample temperature to stabilize in order to correct for changes due to thermal expansion and to ensure that the sample was in thermal equilibrium before acquiring a spectrum.

\noindent\mathbi{DFT calculations:} The electronic band structure of \RVS{} calculated by VASP~\cite{Kresse1994,Kresse1996,KRESSE199615} after a full relaxation on atomic positions with an atom pair-wise correction method (DFT-D3)~\cite{Grimme2011}. The phonon band structure of \RVS{} is calculated using VASP within the Perdew-Burke-Ernzerhof exchange correlation based on density functional perturbation theory. An equivalent k-point mesh of $9\times9\times6$ is used in the self-consistent calculation, and the cutoff energy for the plane-wave basis is 400~eV. Prior to the phonon spectra calculation, crystal structure is relaxed with the residual force on each atom is less than 0.001 eV/\AA.

\section{Commensurability effect}

\begin{figure*}[tb]
\includegraphics[width=14 cm]{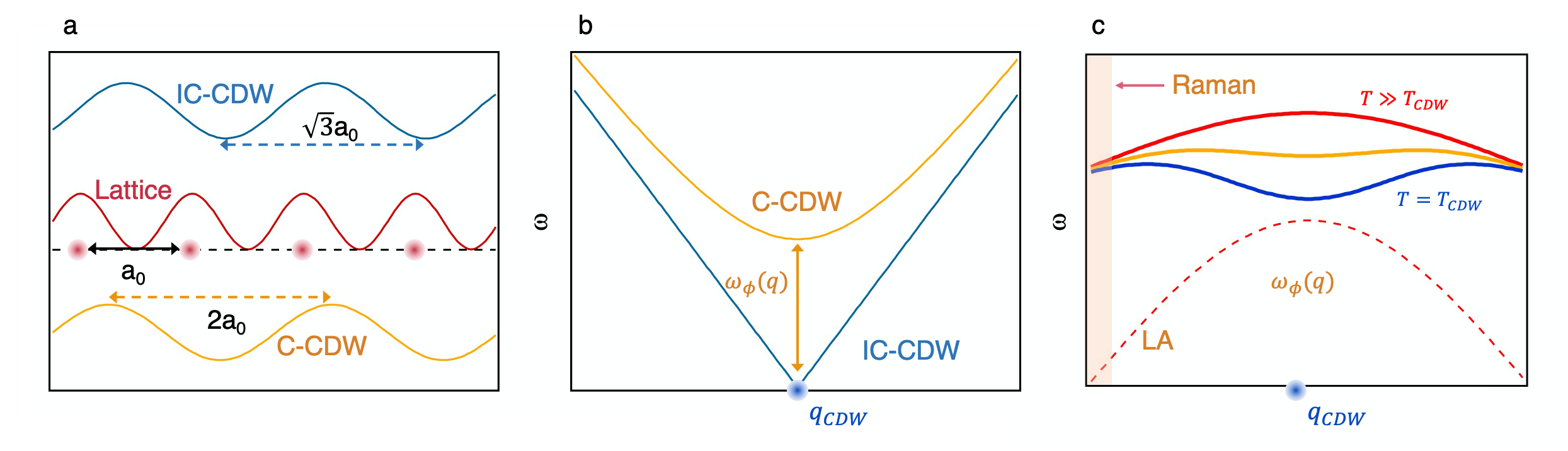}
\renewcommand{\thefigure}{A\arabic{figure}}
\caption{(a) schematically shows commensurate (yellow) and incommensurate (cyan) CDW periods with respect to the lattice (red). (b) Yellow and cyan curves are the calculated phase modes for C-CDW and IC-CDW, respectively. (c) shows the CDW mechanism without electron-acoustic phonon interaction. The temperature dependent particle-hole excitations are shown as red, yellow and blue curves. The dashed line show LA mode, whose energy at $q_{CDW}$ is less than $\omega_{\phi}$. Shaded pink area, indicates the momentum position of Raman measurement. 
\label{FigA2}}
\end{figure*}

For incommensurate CDW (IC-CDW), the condensation energy, $E_{cond}$ is independent of the phase of CDW, $\phi$. For commensurate CDW (C-CDW), however, the condensation energy becomes phase dependent:
\begin{equation}
E_{\text{cond}}(\phi)=-\frac{n(\epsilon_{\text{F}})\Delta_{\text{CDW}}^{2}}{\lambda}(\frac{\Delta}{D})^{M-2}cos(M\phi)
\label{cond}
\end{equation}
where $\lambda$ is the dimensionless EPC constant, $n(\epsilon_{\text{F}})$ and $\Delta$ are the density-of-state at $E_F$ and the CDW gap, respectively. $M$ is the CDW period relative to the lattice constant $a_0$. $D$ is the electronic band width. This phase dependent $E_{\text{cond}}(\phi)$ means that gliding the CDW requires finite energy, corresponding to a finite gap in the CDW phase mode:
\begin{equation}
\omega_{\phi}(q)=(\frac{\omega_{F}^{2}}{M}+v_{F}^2\frac{m}{m^{*}}q^2)^{1/2}
\label{phase}
\end{equation}
where $\omega_{F}^{2}=\frac{4M^{2}}{\lambda}\frac{m\Delta^{2}}{m^{*}}(\frac{\Delta}{D})^{M-2}$. $m^{*}$ is the effective mass. At $q=q_{CDW}$, Eq.~\ref{phase} gives:

\begin{equation}
\omega_{\phi}=\sqrt{\frac{4M}{\lambda}\frac{m}{m^{*}}}\Delta(\frac{\Delta}{D})^{\frac{M-2}{2}}
\label{gap}
\end{equation}
Usually $D\gg\Delta$ and therefore, $\omega_{\phi}\sim0$. However, for $M=2$, which is the case of \AVS{}, $\omega_{\phi}=\sqrt{\frac{4M}{\lambda}\frac{m}{m^{*}}}\Delta$. Most interestingly, a simple parabolic fitting of the ARPES data shown in Fig.~\ref{Fig4} gives $\frac{m}{m^{*}}\sim$3 for the shallow electron and hole pockets, suggesting large $\omega_{\phi}$ in \AVS{}. As we shown in Fig.~\ref{FigA2}c, due to the large phason gap, the particle-hole condensation will not interacting with acoustic phonon modes, as these modes are laying inside the phason gap.

%
\bibliography{ref}
\end{document}